\documentclass[a4paper,11pt]{article}
\usepackage{aaskaiid}
\usepackage{orcidlink}

\newcommand{\kms}{\textrm{km s}^{-1}}
\newcommand{\hi}{\text{H\,\sc{i}}}
\newcommand{\msol}{\textrm{M}_{\odot}}
\newcommand{\pc}{\text{pc}}

\title{The Bright Future of the Dark and Dim Universe}
\ShortTitle{Dark and Dim}

\author[1]{N. Deg\orcidlink{0000-0003-3523-7633}}
\ShortName{Deg et al.} % shortened name list for header 
\author[2]{Alejandro Benitez-Llambay\orcidlink{0000-0001-8261-2796}}
\author[3,4]{Elizabeth A. K. Adams\orcidlink{0000-0002-9798-5111}}
\author[5,6]{N. Chamba \orcidlink{000-0002-1598-5995}}
\author[7]{V. Kilborn\orcidlink{0000-0003-3636-4474}}
\author[8,9]{Filippo M. Maccagni\orcidlink{0000-0002-9930-1844}}
\author[7,10,11]{T. O'Beirne\orcidlink{0009-0007-8779-1827}}
\author[1]{K. Spekkens\orcidlink{0000-0002-0956-7949}}

\affiliation[1]{Department of Physics, Engineering Physics, and Astronomy,Queen’s University,
Kingston ON K7L 3N6, Canada}
\emailAdd{nathan.j.deg@gmail.com}

\affiliation[2]{Dipartimento di Fisica G. Occhialini, Universit\`a degli Studi di Milano Bicocca, Piazza della Scienza, 3 I-20126 Milano MI, Italy}
\affiliation[3]{ASTRON, Netherlands Institute for Radio Astronomy, Oude Hoogeveensedijk 4, 7991 PD Dwingeloo, The Netherlands}

\affiliation[4]{Kapteyn Astronomical Institute, University of Groningen, Postbus 800, 9700 AV Groningen, The Netherlands}
\affiliation[5]{NASA Ames Research Center, Moffett Field, CA 94035, USA}
\affiliation[6]{Department of Space, Earth and Environment, Chalmers University of Technology, Chalmersplatsen 4, Gothenburg SE-412 96, Sweden}
\affiliation[7]{Centre for Astrophysics and Supercomputing, Swinburne University of Technology, Hawthorn, Victoria 3122, Australia}

\affiliation[8]{INAF -- Osservatorio Astronomico di Cagliari, via della Scienza 5, 09047, Selargius (CA), Italy}

\affiliation[9]{Wits Centre for Astrophysics, School of Physics, University of the Witwatersrand, 1 Jan Smuts Avenue, 2000, Johannesburg, South Africa}
\affiliation[10]{European Southern Observatory, Karl-Schwarzschildstrasse 2, D-85748 Garching bei München, Germany}
\affiliation[11]{CSIRO Space \& Astronomy, PO Box 1130, Bentley WA 6102, Australia}
\abstract{This chapter investigates the low-mass frontier of galaxy formation through two complementary populations: the starless Reionization-Limited \hi\ Clouds (RELHICs) that trace the ``dark'' Universe, and the faint, gas-rich galaxies that define the ``dim'' Universe. RELHICs offer pristine laboratories for probing the distribution of DM on sub-galactic scales, providing a direct test of the Lambda Cold Dark Matter ($\Lambda$CDM) model predictions. The dim Universe provides statistical constraints on cosmology, galaxy formation and evolution, as well as baryoninc physics through key observables including the low-mass end of the neutral-hydrogen mass function (HIMF), the neutral-hydrogen velocity function (HIVF), and the low-mass end of the baryonic Tully-Fisher relation (bTFR). This chapter outlines core science questions that can be tackled leveraging radio observations of both the dark and dim Universe.  Additionally, it outlines strategies to identify RELHICs amid tidal or pressure-confined contaminants, while providing observational predictions for the dim Universe.
The Square Kilometre Array (SKA) in its mid-frequency Array Assembly 4 (AA4) configuration will, for the first time, resolve the internal gas structure of nearby RELHICs and build deep, wide-area datasets that definitively constrain the HIMF, HIVF, and bTFR down to masses of $10^{6}~\msol$---offering a complete observational framework to test the $\Lambda$CDM paradigm and the baryonic processes that shape the faint end of galaxy formation.}

\begin{document}
\maketitle
\tableofcontents

\section{Introduction}
\label{sec:Introduction}

The Lambda-Cold Dark Matter ($\Lambda$CDM) model has emerged as the standard paradigm of modern cosmology, providing a remarkably successful framework for describing the formation and evolution of structure in the Universe. A cornerstone of this model is the bottom-up growth of structure, in which the smallest dark matter (DM) halos collapse first at high redshift and subsequently merge to form the more massive systems that host galaxies today~\citep[e.g.,][]{Blumenthal1984}. Present-day low-mass halos are therefore the residuals of this hierarchical process and serve, to some extent, as analogues of the early building blocks of galaxies. Understanding their properties is essential for constraining both the physics of galaxy formation and the underlying cosmological model.   Observationally, studying low mass halos means studying low mass galaxies.  At low masses, \hi\ is the dominant baryonic component of star forming galaxies.  Because the size and scale of the \hi\ reservoirs are sensitive to both internal and environmental processes, the \hi\ properties of low mass, star forming galaxies are particularly valuable cosmological probes.

A fundamental prediction of hierarchical structure formation is a halo mass function (HMF) that rises steeply toward lower masses, scaling as $dN/dM \propto M^{-1.9}$~\citep{Press1974, Jenkins2001}. The Universe should therefore be overwhelmingly populated by low-mass DM halos, which far outnumber their massive counterparts. However, observations of the galaxy stellar mass function~\citep[GSMF; e.g.,][]{Driver2022} and the neutral hydrogen (\hi) mass function~\citep[HIMF; e.g.,][]{Jones2018, Ma2025} reveal a substantially flatter slope at the low-mass end. 

These fundamental discrepancies imply that baryonic processes must strongly suppress the efficiency of gas accretion and galaxy formation within shallow potential wells~\citep[see, e.g., the review by][]{Bullock2017}. This suppression sculpts the low-mass frontier into two distinct but physically-linked populations: a vast number of halos rendered entirely starless (the `dark' universe) and a smaller but crucial population of faint, gas-rich systems that sit at the very threshold of star formation (the `dim' universe).  

Within $\Lambda$CDM, this inefficiency of gas accretion and galaxy formation in shallow potentials arises naturally from the combined action of stellar feedback and cosmic reionization. Supernova-driven winds, stellar radiation, and heating from the ultraviolet background (UVB) inhibit gas cooling and accretion. During the epoch of reionization, the UVB further heated the intergalactic medium, preventing gas infall or even photoevaporating existing reservoirs~\citep[e.g.][and references therein]{Thoul1996, Quinn1996}. These effects established a redshift-dependent critical halo mass, $M_{\rm crit}(z)$, below which galaxy formation is supressed~\citep[e.g.,][and references therein]{Gnedin2000, Hoeft2006, Benitez-Llambay2020}. Consequently, a vast population of halos is predicted to remain either extremely faint or entirely `dark' today~\citep[e.g.,][]{Sawala2016}, with some of them containing only primordial hydrogen gas in hydrostatic equilibrium with their DM potentials, but  no stars. The search for these starless DM-dominated Reionization-Limited \hi\ Clouds~\citep[RELHICs;][] {Benitez-Llambay2017} represents a key test of the hierarchical nature of structure formation.

Beyond simply confirming their existence, these pristine, starless systems provide a unique laboratory for tackling another fundamental small-scale challenge: the internal structure of DM halos. While collisionless simulations predict steep, cuspy central density profiles~\citep{Navarro1997}, observations of low-mass galaxies reveal a significant diversity in rotation-curve shapes, with many systems exhibiting flatter, constant-density cores~\citep[e.g.][]{Flores1994, Moore1994, Oh2015, Oman2015}. A central question is whether this diversity signals new DM physics---such as Warm or Self-Interacting DM, which naturally produce cores~\citep[e.g.,][]{Bode2001, Spergel2000}---or reflects the dynamical impact of baryonic processes like repeated, supernova-driven outflows, which have been shown in high-resolution simulations to transform cusps into cores~\citep[e.g.,][and references therein]{Governato2010, Pontzen2012, DiCintio2014, Benitez-Llambay2019}. Furthermore, the inference of inner density slopes is often complicated by systematic effects and modeling choices, including beam smearing, inclination uncertainties, and non-circular motions~\citep[e.g.,][]{vandenBosch2000, Oman2019a, Pineda2017}. Because they lack the stars necessary to drive such outflows, and their gas is non-self-gravitating, RELHICs should preserve the primordial shape of their host DM halos, offering a uniquely clean probe of the pristine density profile predicted by $\Lambda$CDM, and a powerful test to distinguish between baryonic and fundamental physics solutions.

Whereas starless halos reveal the pristine structure of DM, faint, gas-rich galaxies near the threshold of star formation trace how baryonic processes reshape the low-mass Universe, providing key tests of cosmology and galaxy evolution.
These systems encode the physical transition from the starless to the star-forming regime. Their shallow potentials exacerbate the effects of baryonic processes, whereas their low baryon content makes them ideal laboratories for statistically probing the structure of DM halos. Ultimately, their gas content, kinematics, and scaling relations---traced through the HIMF, the \hi\ velocity function~\citep[HIVF; e.g.,][]{Zavala2009, Zwaan2010,Papstergis2015}, 
and the baryonic Tully–Fisher relation~\citep[bTFR; e.g.][]{McGaugh2000}---offer powerful constraints on the interplay between baryons and DM and on the influence of environment at the low-mass frontier. Together, these metrics serve as statistical probes of how baryonic physics reshapes the low-mass end of structure formation.

Together, the dark and dim Universe mark the threshold where cosmology meets baryonic physics. A comprehensive understanding of these issues thus requires a dual approach: first, the search for the `dark' Universe, embodied by primordial, starless halos (RELHICs) that represent the `missing' halo population predicted by theory; and second, the study of the `dim' Universe, where the faintest observable galaxies allow us to statistically probe the physical processes, either internal or environmental, that sculpt structure at this frontier.

This chapter explores the key science questions arising from this framework. To this end, Section~\ref{sec:CoreScience} is structured to address both frontiers. We begin with the dark Universe, examining the observational search for primordial RELHICs and tidally-formed dark systems. We then turn to the dim Universe, exploring key core questions probed by statistical samples of faint galaxies, including the HIMF, the HIVF, and the bTFR.  We end by discussing how internal processes and the external environment shape these low-mass systems.
%probing the faint galaxy population through key statistical measures—--the HIMF and HIVF—--and scaling relations like the bTFR. Finally, we investigate how internal properties and external environment shape these low-mass systems. 
Section~\ref{sec:Prospects} outlines how the Square Kilometre Array (SKA), in synergy with other observatories, will illuminate both of these frontiers. Finally, Section~\ref{sec:Summary} presents a summary of this chapter and the outlook for studies of the dark and dim Universe in the near future.

\section{Key Science Questions}
\label{sec:CoreScience}

As discussed above, understanding the low-mass frontier of cosmic structure formation requires tackling two distinct but complementary populations: the truly primordial starless ``dark'' halos (RELHICs), and the faint, gas-bearing ``dim'' galaxies that sit at the threshold of star formation. This section details key scientific questions that can be addressed by the SKA for both dark and dim galaxies.

\subsection{The Search for the Dark Universe}

%\subsubsection{Primordial Dark Galaxies: Reionization Limited \hi\ Clouds}
\label{Sec:RELHICs}

The definitive confirmation of a cosmologically significant population of starless halos would represent a triumph for the $\Lambda$CDM model. But does the vast population of low-mass, starless halos predicted by $\Lambda$CDM actually exist? The search for these objects relies on untargetted, wide-area \hi\ surveys, a challenging endevour given that RELHICs are small, intrinsically faint, and can be confused with other \hi\ structures such as tidal debris. Key predictions that distinguish them are their roughly spherical shape and distinct kinematic signature: a narrow \hi\ emission line consistent with thermal broadening ($T\approx 2\times 10^{4}$ K) rather than the broader profiles indicative of rotation or tidal forces~\citep[see, e.g.,][]{Benitez-Llambay2017}. 

Closely related is the question of what physical processes establish the lower mass limit for galaxy formation. RELHICs are understood to occupy halos just below the critical mass threshold where gas can collapse, and thus star formation becomes impossible due to the photoheating of the cosmic UVB~\citep{Benitez-Llambay2020}. Identifying a population of these systems and comparing their properties with those of star-forming dwarf galaxies would allow us to measure this transitionary mass scale directly~\citep[e.g.][]{Benitez-Llambay2021, Pereira-Wilson2023}. This provides a powerful empirical test for theories of galaxy formation suppression, allowing us to quantify the relative impact of the UVB versus environmental mechanisms like ram-pressure stripping that can also remove a halo's gas reservoir~\citep[e.g.][and references therein]{Benitez-Llambay2013, Herzog2023, Pasha2023, Thompson2023}.

On a complementary front, these pristine systems offer an unparalleled opportunity to address one of the most significant small-scale challenges: What is the primordial inner density structure of DM halos? In luminous galaxies, the question of whether halos are cuspy (as predicted by collisionless DM simulations) or cored (as inferred by observations of some nearby dwarfs) is complicated by baryonic feedback, which can dynamically alter the central potential and reshape the inner DM content of galaxies~\citep[e.g.,][and references therein]{Pontzen2012, DiCintio2014, Tollet2016, Benitez-Llambay2019}. Because they are starless, RELHICs are pristine laboratories where the DM halo has remained undisturbed. Cosmological simulations predict their gas content should be in a nearly spherical distribution and held in hydrostatic and thermal equilibrium within the DM potential. This equilibrium links the observable gas structure directly to the underlying DM profile, offering a uniquely clean test of the fundamental nature of DM---free from the complexities of galaxy formation physics.

However, the reliability of this test depends on several physical assumptions regarding the state of the neutral gas. Specifically, the inference of a DM profile from the H I distribution assumes that the gas is in hydrostatic and thermal equilibrium within the halo potential, with a velocity dispersion dominated by thermal broadening and negligible contributions from turbulence or bulk non-circular motions. Furthermore, it assumes that the systems are truly isolated and are not subject to significant external pressure confinement from a surrounding hot medium, which could otherwise mimic the signatures of a DM halo. Distinguishing between these scenarios remains a significant challenge, as departures from equilibrium or the presence of non-thermal support could bias the inferred inner DM distribution. One of the primary goals for the SKA at AA4 is to explicitly test these assumptions by providing the resolution and sensitivity needed to map the internal gas temperature and kinematics, thereby determining whether the gas is indeed a faithful tracer of the underlying dark matter distribution.

The recent candidate `Cloud-9' provides the first compelling evidence addressing these issues. Observations with the Five-Hundred-Meter Spherical Telescope (FAST) revealed a dynamically cold system with a narrow velocity width ($w_{50} \approx 12$ km/s) consistent with a pressure-supported, non-rotating massive gas cloud ($M_{\rm HI} \approx 10^6 \rm \ M_{\odot}$), embedded in a massive DM halo of mass remarkably close to $M_{\rm crit}(z=0)$~\citep{Zhou2023, Benitez-Llambay2023, Karunakaran2024}. Crucially, recent deep follow-up imaging with the Hubble Space Telescope provided the decisive evidence, revealing no discernible stellar population and placing a stringent upper limit on its stellar mass of less than $10^{3.5} \rm \ M_{\odot}$~\citep{Anand2025}. While the gas distribution of Cloud-9 is in tension with $\Lambda$CDM if interpreted as an average RELHIC, possibly signaling the presence of a large DM core in its center~\citep[][]{Benitez-Llambay2023}, subsequent VLA and GBT observations suggest instead that the system is likely experiencing interactions with M94~\citep{Karunakaran2024,Benitez-Llambay2024}, further complicating its interpretation and raising the possibility that the system is not primodial in nature.
    
Despite the promise of Cloud-9, a major challenge remains: unambiguously establishing whether such systems are truly isolated, self-gravitating DM halos or instead the byproducts of other astrophysical processes. Confounding sources abound. Some high-velocity clouds~\citep[][]{Wakker1997}, ultra-compact high velocity clouds~\citep[UCHVC; e.g.,][]{Adams2013}, or extended tidal debris from galaxy mergers and interactions~\citep[e.g.][]{Hibbard2001, Putman2003, Nidever2010} may mimic some of the observed characteristics of RELHIC candidates, including compact, cold \hi\ morphologies and narrow line widths~\citep[][]{Benitez-Llambay2017}.

Likewise, small \hi\ clouds may also exist in pressure equilibrium within the hot circumgalactic medium (CGM) of massive galaxies, giving rise to quasi-stable configurations that do not require DM~\citep[e.g.][and references therein]{Marinacci2010, Armillotta2017}. The SECCO 1 system in the Virgo Cluster \citep{Bellazzini2018}, one system within a class of young, blue, isolated stellar features in Virgo with high metallicities and young stellar populations posited to be remnants of ram pressure stripping events \citep{Jones2022,Bellazzini2022,Dey2025G}, exemplifies this ambiguity.

The SECCO~1 system in the Virgo Cluster exemplifies this ambiguity. The SECCO survey, designed to identify stellar counterparts to compact \hi\ clouds, discovered instead a low-mass, star-forming object whose gas is likely confined by the high ambient pressure of the intracluster medium rather than by its own DM halo~\citep{Bellazzini2018}. This case underscores a critical point: \hi\ clouds in dense environments can mimic many of the observational properties expected of starless halos.

A second major source of confusion arises from tidal dwarfs and tidally stripped \hi\ clouds, which can appear strikingly similar to dark galaxies in both morphology and kinematics. Unlike RELHICs, however, these systems are baryon-dominated and form from gas---and occasionally stars---removed from massive galaxies through tidal interactions or ram-pressure stripping~\citep[e.g.,][]{Kaviraj2012, Lee-Waddell2014, OBeirne2024, Taylor2022, Oosterloo2005}. While some examples show unmistakable interaction signatures, others occupy more ambiguous regimes where the distinction between a tidal origin and a primordial halo remains uncertain.

The case of VIRGOHI21 also illustrates this complexity. Initially identified as a dark galaxy candidate~\citep{Davies2004, Minchin2005}, its apparent isolation and broad velocity gradient initially seemed inconsistent with tidal debris. Subsequent hydrodynamical simulations~\citep{Bekki2005} and deep environmental studies~\citep{Haynes2007, Duc2008}, however, demonstrated that tidal features can reproduce such characteristics—exhibiting wide velocity spreads, low surface brightness, and apparent separation from their progenitors. VIRGOHI21 is now understood as part of an extended \hi\ tail produced by repeated high-speed encounters within the Virgo Cluster, rather than as an isolated, dark system.

In gas-rich and dynamically complex environments---such as compact groups or merging systems---this ambiguity becomes especially acute. High-resolution imaging is often essential to reveal faint stellar components and distinguish genuine dwarf galaxies from transient tidal debris. Stephan’s Quintet~\citep{Amram2002, Williams2002} and the M81 group~\citep{Boyce2001, Gozman2024} provide instructive examples: increasingly sensitive observations have uncovered both starless \hi\ clouds and tidal dwarfs with young stellar populations, illustrating the continuum of possible outcomes in interacting environments.

For large-scale surveys, however, detailed simulation-based analysis of each candidate is not feasible. Instead, statistical approaches are being developed to separate tidal debris from isolated, self-gravitating systems. For example, \citet{Kwon2025} applied an environment-based distance criterion within a $\pm400$~km~s$^{-1}$ velocity window in the ALFALFA survey, while \citet{OBeirne2025} used virial-radius and velocity-based cuts in WALLABY to identify sources without obvious tidal origins. These methods offer scalable ways to flag likely contaminants, but the challenge remains formidable. As the sensitivity and volume of next-generation facilities such as the SKA expand, disentangling pressure-confined and tidal systems from truly isolated, dark-matter–dominated halos will be critical for robustly mapping the dark low-mass frontier. It is only in these isolated environments where the RELHIC hypothesis for a specific galaxy can be tested as it requires all baryonic alternatives to be ruled out.

Achieving these goals demands a synergy between untargetted and targeted surveys that combine sensitivity, volume, and resolution. While wide-area, untargetted 21-cm surveys such as those conducted with ALFALFA and FAST have demonstrated the feasibility of uncovering compact, starless \hi\ clouds, their limited depth restricts detections largely to the nearby Universe~\citep[e.g.][]{Adams2013, Haynes2018, Zhang2024}. This limitation is further compounded by the intrinsically low abundance of such systems: as shown in Figure~\ref{fig:HIMF}, the expected number density of RELHICs is lower than that of field galaxies by roughly a factor of five at $M_{\rm HI} \approx 10^{6},\msol$ (see Section ~\ref{Sec:HIMassFunction} for a full discussion).

Next-generation facilities---most notably the SKA-Mid at AA4---will provide the transformative leap required. Current surveys are limited by a combination of sensitivity, which restricts them to the nearby Universe, and survey speed, which makes mapping vast, truly isolated cosmic voids impractical. By overcoming these limitations, the SKA's AA4 instrument will build the statistically robust sample required to distinguish between tidal debris, pressure-confined clouds, and genuinely \hi-rich starless halos, allowing the existence of RELHICs to be confirmed beyond reasonable doubt.

%An inevitable outcome of these searches will be the discovery of extremely faint, gas-bearing galaxies at the threshold of star formation. At the mass scale of RELHICs, most halos are expected to have already formed at least a minimal stellar component prior to reionization~\citep[e.g.,][and refereces therein]{Sawala2016, Fitts2017, Benitez-Llambay2020}, and indeed, systems such as Leo~P~\citep{Giovanelli2013} and KK~153~\citep{Xu2025}—first identified in 21-cm emission—exemplify this transitional regime. Thus, the effort to uncover truly starless halos will, as a byproduct, reveal the faintest galaxies capable of forming stars, providing a crucial observational link to the ``dim'' population that is best studied through the statistical methods discussed next.

\subsection{The Dim Universe: A Statistical Approach to Cosmological Tests}
\label{sec:DimUniverse}

Studies of the dim universe are complementary to those of the dark universe, probing different physics and testing different questions.  Unlike dark galaxies, which require a 'needle-in-a-haystack' approach, the population of low mass and low surface brightness galaxies is significant, enabling statistical studies of cosmology.
The core predictions of the $\Lambda$CDM paradigm are inherently statistical: the HIMF, the HIVF, and the bTFR.  And beyond cosmological predictions, studying the dim universe explains how the baryons interact with DM, and how environment reshapes and drives galaxy evolution. 

\subsubsection{The \hi\ Mass Function}
\label{Sec:HIMassFunction}

The observed HIMF is substantially flatter at the low mass end than predictions of the HMF from $\Lambda$CDM.  This difference effectively quantifies the degree to which baryonic processes must suppress or remove gas from the vast majority of low-mass halos. The precise shape of the HIMF, particularly its faint-end slope, therefore encodes the efficiency of galaxy formation and the impact of feedback and environmental stripping across cosmic time. The HIMF is typically described by a Schechter function~\citep{Schechter1976}:
\begin{equation}
    \phi(M_{\rm HI}) = \frac{dn}{d\log_{10} (M_{\rm HI})} = \ln(10)\phi_\star\bigg(\frac{M_{\rm HI}}{M_\star}\bigg)^{\alpha+1}\exp\bigg(-\frac{M_{\rm HI}}{M_\star}\bigg)
\end{equation}
where $\phi$ is the galaxy number density as a function of mass, $\alpha$ denotes the faint-end slope, $M_\star$ marks the characteristic “knee” mass, and $\phi_\star$ is a normalization constant. The shape of the HIMF encodes the efficiency with which galaxies acquire, retain, and lose their cold gas reservoirs. By measuring the HIMF across cosmic time and in different environments, one can trace how neutral gas fuels star formation, quantify how feedback and stripping regulate this supply, and test cosmological models that predict the abundance and evolution of low-mass halos.

Untargetted \hi\ surveys such as HIPASS~\citep[e.g.,][]{Meyer2004} and ALFALFA~\citep{Giovanelli2013} provided the first robust determinations of the local HIMF. ALFALFA, in particular, detected large numbers of galaxies ($\ge10^{2}$ per mass bin) down to $M_{\rm HI}=10^{7.5}~\msol$, as well as fewer detections (a total of 144) down to $10^{6}~\msol$.  Based on these, \citet{Jones2018} established a faint-end slope $\alpha \approx -1.3$ in the field.  However, the faint end, (i.e. below $10^{7.5}~\msol$) remains poorly constrained with large uncertainties.

\begin{figure}[h]
    \centering
	\includegraphics[width=0.9\columnwidth]{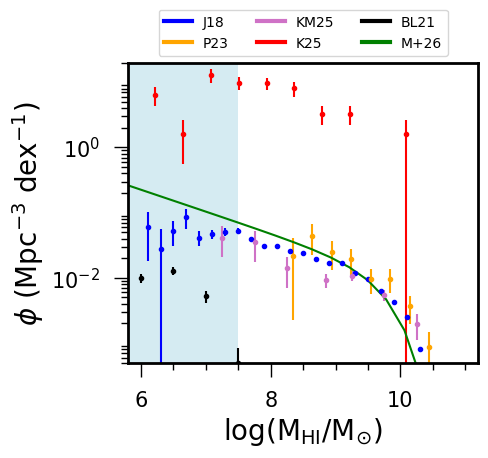}
    \caption{The measured HIMF from ALFALFA (blue; \citealt{Jones2018}), MIGHTEE-HI (orange; \citealt{Ponomareva2023}), LADUMA (light purple; \citealt{Kazemi-Moridani2025}), and Fornax (red; \citealt{Kleiner2025}).  Also shown is the predicted HIMF for RELHICS (black), estimated from the cosmological simulations presented by~\citet{Benitez-Llambay2021}.  The green line shows that \citet{mayor_2026} HIMF fit to the GAEA model \citep{DeLucia2024}.  The shaded turquoise region goes from $\log_{10}(M_{\hi}/\msol)=7.5$ to low-masses and indicates both the upper limit on RELHIC masses as well as the region where most surveys have low galaxy counts.} 
    \label{fig:HIMF}
\end{figure}

The advent of SKA precursors and pathfinders—--including MIGHTEE~\citep[$z<0.083$;][]{Ponomareva2023}, LADUMA~\citep[$z<0.088$;][]{Kazemi-Moridani2025}, FAST~\citep[$z<0.05$;][]{Ma2025}, and WALLABY—--is now transforming this field. These surveys are delivering the first representative, volume-limited catalogs needed to measure the HIMF across different environments and redshifts. 
Figure \ref{fig:HIMF} shows the HIMF measurements of ALFALFA compared to those from MIGHTEE-HI and LADUMA, as well as a theoretical HIMF from simulations \citep{mayor_2026}.  These representative field samples show significant variation at the low-mass end, highlighting the need for more observations below $10^{7.5}~\msol$.  This low mass HIMF is also important to simulations, as it is where various sub-grid physics and feedback processes make it challenging to match precisely.  Improved measurements of the HIMF will in turn be able to constrain simulations, with the caveat that it will be important to take into account diffeing measurement techniques between simulations and observations.  Excitingly, there are already recent improvements in the measurement of the low-mass end coming from FAST.  For example, \citet{Ma2025} combined observations from HIPASS and ALFALFA with FAST to construct an updated HIMF that suggests a double Schecter fit provides a better description than a single fit.  However, deeper surveys will be needed to confirm this result.

It has long been understood that the HIMF also varies as a function of environment. Early, environment-focused studies in clusters~\citep{Verheijen2000,Busekool2021,Yu2025} and groups~\citep{Pisano2011,Kilborn2009,Westmeier2017,Jones2020} have suggested that the HIMF slope is flatter in denser environments.  These results are consistent with some studies that subdivide larger samples by local density \citep{Zwaan2005}, but there are studies that find differing results \citep{Springob2005, Said2019}.  Additionally, studies of void galaxies have revealed similar slopes in the low mass end, but with offsets in the total number of galaxies \citep{Moorman2014}.  A striking recent result comes from the MeerKAT Fornax Survey~\citep{Serra2023}, which achieved unprecedented sensitivity and completeness down to $M_{\rm HI} \sim 10^6\,M_\odot$ \citep{Kleiner2025}. As shown in Figure \ref{fig:HIMF}, the resulting HIMF exhibits a pronounced cutoff below $M_{\rm HI} \sim 10^{7}\,M_\odot$: where $\sim$30–40 detections were expected from extrapolating the field relation, only five were observed. This represents the first direct evidence for a low-mass truncation in a dense cluster environment, demonstrating the extreme efficiency of \hi\ removal mechanisms—likely a combination of ram-pressure stripping and tidal heating. Comparable studies in looser groups show flatter slopes rather than cutoffs, while observations in rich clusters such as Coma reveal a bimodal \hi\ distribution at higher masses ($M_{\rm HI} \sim 10^{8}\,M_\odot$), reflecting the coexistence of gas-rich and strongly depleted galaxies~\citep{healy2021,molnar2022}.

%Integrating the HIMF yields the cosmic \hi\ mass density, $\Omega_{\rm HI}$, a critical global measure of neutral gas evolution. Interestingly, current measurements indicate remarkably little evolution in $\Omega_{\rm HI}$ over cosmic time, in stark contrast to the strong evolution observed in the molecular gas phase~\citep{Walter2020}. The MIGHTEE survey has extended HIMF measurements to $z\sim0.084$ without finding significant change~\citep{Ponomareva2023}, although its sensitivity threshold of $M_{\rm HI} \gtrsim 10^{8}\,M_\odot$ precludes a direct comparison with deeper local surveys. A consistent determination of the low-mass HIMF—and thus of $\Omega_{\rm HI}$—remains a central missing piece in our picture of how galaxies build and exhaust their cold gas reservoirs.

Resolving these questions—--specifically, the precise faint-end slope of the HIMF and its environmental dependence
%, and its redshift evolution
---requires sensitivity to $M_{\rm HI} \lesssim 10^{7}~M_\odot$ across diverse environments and cosmic epochs. This is precisely the regime where the dark and dim Universes overlap. While MeerKAT has achieved such sensitivies out to ~20 Mpc, this requires $\sim55$ hr of integration time \citep{deBlok2024,Kurapati2025}.  The SKA-Mid will be transformative in this regime.  With detection limits reaching $M_{\rm HI} \sim 10^{6}\,M_\odot$ out to distances of $\ge50$ Mpc  in only a few hours per square degree (see Sec. \ref{sec:Prospects} for quantitative predictions), it will enable direct measurement of the HIMF cutoff, map its variation with environment, and trace the evolution of $\Omega_{\rm HI}$ across cosmic time. Reaching these depths will also allow resolved \hi\ kinematic studies of dwarf galaxies, constraining their gas removal timescales~\citep{Kleiner2021}, baryon-to-DM ratios~\citep{Maccagni2024b}, and the efficiency of the baryon cycle~\citep[e.g.,][]{ManceraPina2025}. Together, these advances will bridge the gap between the dark and luminous low-mass populations, illuminating the physical processes that govern the transition between the two.

\subsubsection{The \hi\ Velocity Function}
\label{Sec:HIVelFunction}

Complementing the low-mass end of the HIMF, the \hi\ Velocity Function (HIVF) probes the distribution of dynamical masses on a population level. The HIVF measures the number density of systems (either galaxies or halos) as a function of their rotation speed, which is more closely linked to the inner mass distribution of the host DM halo for galaxies (and the temperature of the gas for RELHICs) than the \hi\ mass itself. Observationally, studies of the HIVF typically use inclination-corrected spectral line widths---such as $w_{50}$ or $w_{20}$, the width at 50\% or 20\% of the peak flux, respectively---as a proxy for the maximum rotation velocity~\citep{Zavala2009,Zwaan2010,Papstergis2015}.

Since its first measurements, the HIVF has presented a significant challenge to the $\Lambda$CDM model. Straightforward interpretations of cosmological simulations predict a steep rise in the number of low-mass DM halos, which should translate to a correspondingly large population of low-velocity galaxies. However, observations consistently find far fewer low-rotation galaxies than predicted, a tension that has been confirmed by multiple studies~\citep{Zavala2009,Zwaan2010,Papstergis2015,Klypin2015}. This discrepancy is most pronounced at rotation velocities below $\sim$50 km/s, where simulations overpredict the abundance of galaxies by an order of magnitude or more.

Significant work has been undertaken to understand the origin of this discrepancy. From a theoretical perspective, more nuanced analyses that account for baryonic processes are beginning to bridge the gap. Hydrodynamical simulations show that feedback from supernovae can expel gas, reducing the growth rate of halos and altering the relationship between a galaxy's observed \hi\ line width and the circular velocity of its host halo~\citep[e.g.][]{Sawala2015, Schaller2015}. These effects are complex and non-linear; for instance, baryonic physics tends to cause \hi\ line widths to underestimate the true halo velocity in low-mass systems. From an observational standpoint, researchers have carefully investigated systematic biases, survey selection effects, and the large scatter in the relationship between \hi\ mass and circular velocity, all of which complicate a direct comparison with theoretical predictions~\citep{Maccio2016,Brooks2017,Chauhan2019,Dutton2019,Oman2022}.

A crucial step forward is to move beyond simple profile widths and towards the use of resolved kinematic models for large, statistically complete samples of galaxies. Full kinematic modeling provides more reliable inclinations and delivers the rotation velocity as a function of radius, allowing for more robust measurements of the underlying gravitational potential, including corrections for pressure support (asymmetric drift)~\citep{Iorio_2016}. To date, however, samples with full kinematic models have been limited to a few hundred galaxies from targeted surveys, which are often biased towards `well-behaved' systems and lack the statistical power to robustly constrain the low-velocity end of the HIVF~\citep{Begum2008b,Lelli2016b,Koribalski2018,Ponomareva2021,Deg2022,Murugeshan2024}. Moreover, the low rotation regime is the most challenging to model \citep{Deg2025}, requring a careful understanding of systematics and consistent rotation definitions (see Sec. \ref{Sec:bTFR}).  And, the pressure support contributions are more significant, meaning that a careful treatment is necessary.  Despite these challenges, the WALLABY survey is poised to make a significant advance by generating kinematic models for thousands of galaxies, though selection effects may still bias its sample toward systems with larger rotation velocities~\citep[][]{Deg2025}.

Resolving the HIVF tension unambiguously requires building statistically significant samples of kinematically modeled galaxies extending to the lowest rotation speeds ($V_{\rm rot} \lesssim 20$ km~s$^{-1}$). The critical barrier is observational: obtaining resolved kinematic maps for thousands of galaxies with %faint, low-surface-brightness 
\hi\ disks ($V_{\rm rot} \lesssim 20$ km~s$^{-1}$) is beyond the reach of current facilities. This demands a combination of sensitivity, angular resolution, and survey speed that only the SKA-AA4 configuration can provide. With its capabilities, it will be possible to conduct deep, wide-area surveys that can detect and resolve the \hi\ kinematics of thousands of low-mass galaxies across a range of environments. By combining these unprecedented observational samples with increasingly sophisticated mock observations from cosmological simulations~\citep{Oman2019b}, it will be possible to perform a definitive test of the predicted HIVF and robustly constrain the interplay between baryonic physics and DM at the faint end of galaxy formation.

\subsubsection{The Baryonic Tully--Fisher Relation}
\label{Sec:bTFR}

Another key diagnostic for probing the dim Universe is the baryonic Tully--Fisher relation (bTFR), a fundamental scaling law connecting a galaxy’s total baryonic mass ($M_{\mathrm{bar}} = M_* + M_{\mathrm{gas}}$) to the depth of its DM potential well, as traced by its rotation velocity, $V_{\mathrm{circ}}$~\citep{McGaugh2000}. While the HIMF and HIVF characterize the gas content and potential wells of galaxies statistically, the bTFR links them directly at the level of individual systems. It therefore provides a powerful test of the predicted relationship between galaxies and their host halos in $\Lambda$CDM, particularly important at the lowest mass scales~\citep[e.g.,][]{Lelli2019, McQuinn2022}. Two central questions currently frame the ongoing and future research at the faint end of this relation.

\begin{figure}[h]
    \centering
	\includegraphics[width=0.9\columnwidth]{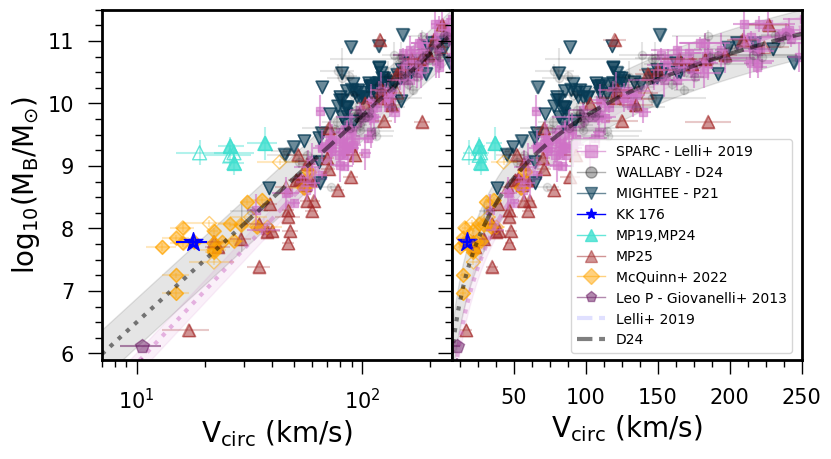}
    \caption{
    The bTFR as calculated from both \citet{Lelli2019} (purple line) and \citet{Deg2024} (grey line). This plot includes the SPARC galaxies, WALLABY galaxies, MIGHTEE-HI measurements \citep{Ponomareva2021}, and the \citet{ManceraPina2025} sample of tRGB distance measured galaxies.  The plot also includes the UDGs from \citet{ManceraPina2019}, the diffuse dwarf KK 176 \citep{Dudley2025}, low-mass galaxies from \citet{McQuinn2022}, and Leo~P from \citet{Giovanelli2013}. 
    Dashed lines show the measured bTFR; dotted lines indicate extrapolations to lower masses.
    }
    \label{fig:bTFR}
\end{figure}

The first question concerns whether the bTFR---nearly a single power law for massive galaxies~\citep[e.g.][and references therein]{Stark2009, Gurovich2010, Lelli2019}---bends or breaks at low masses. The SPARC sample, which extends to baryonic masses as low as $M_{\mathrm{bar}}\sim10^{8}\,\mathrm{M_\odot}$, established a tight, linear relation based on the flat portions of galaxy rotation curves~\citep{Lelli2019}. More recently, however, a sample of lower-mass galaxies compiled by~\citet{McQuinn2022} was found to exhibit a turndown in this relation when the underlying DM distribution is modeled with an inner core. When instead a cuspy DM profile is adopted---arguably a poorer fit to the data---the turnover largely disappears. Regardless of these modeling uncertainties, the result can be interpreted as tentative evidence that the bTFR may indeed bend at the lowest masses, as shown in Figure~\ref{fig:bTFR}. In contrast, new measurements based on kinematic models of WALLABY detections---reaching similar mass scales---agree well with SPARC where they overlap but yield a significantly different extrapolation to lower masses~\citep{Deg2024}. 

This discrepancy highlights a critical challenge: the bTFR is highly sensitive to the definition of rotation velocity adopted and the underlying modelling~\citep[e.g.][]{Ruan2025}. The SPARC sample uses $V_{\mathrm{flat}}$ (the velocity of the flat part of the rotation curve), the WALLABY sample uses $V_{\mathrm{HI}}$ (the velocity at the \hi\ radius), the MIGHTEE-HI \citep{Ponomareva2021} and \citet{ManceraPina2025} measures use the outermost measured rotation point while \citet{McQuinn2022} low-mass sample, along with the \citet{Giovanelli2013} measurement for Leo P employ velocities from position-velocity diagrams. As discussed by~\citet{Lelli2019}, such differences can systematically shift galaxies along the relation. At $M_{\mathrm{bar}}\approx 10^{7}\,\mathrm{M_\odot}$, the offset between the extrapolated relations of \citet{Lelli2019} and \citet{Deg2024} is approximately $\approx 5~\mathrm{km~s^{-1}}$, underscoring the need for both large statistical samples and consistent kinematic definitions.

The second question is whether certain galaxy populations deviate from the bTFR altogether, which would pose a major challenge to our understanding of galaxy formation. This issue is most acute for gas-rich ultra-diffuse galaxies (UDGs)---systems with the stellar mass of dwarfs but the physical size of $M_{\star}$ galaxies~\citep{ManceraPina2019, ManceraPina2020, Hu2023, Du2024}. Several studies have suggested that some UDGs are DM-deficient and lie far from the established bTFR (Figure~\ref{fig:bTFR}), though these claims remain debated~\citep{Brook2021, Sellwood2022, Lelli2024}.%, and in at least one case, deeper data revealed significantly more DM than initially reported~\citep{ManceraPina2022}.

The UDG classification itself is arbitrary, and these systems may simply represent the diffuse tail of the dwarf galaxy population~\citep{Motiwala2025}. They are notoriously difficult to model because of their low surface brightness and slow rotation. However, recent WALLABY observations of the nearby diffuse dwarf KK~176 show that it is DM-dominated and lies directly on the \citet{Deg2024} bTFR~\citep{Dudley2025} (indicated by the blue star in Figure~\ref{fig:bTFR}). This result---consistent with both lower- and higher-mass dwarfs---suggests that at least some diffuse galaxies remain fully consistent with standard cosmological expectations.

Resolving the nature of any potential bend in the bTFR and the true dynamics of gas-rich UDGs will require statistically significant samples of resolved, low-mass galaxies. These are among the most challenging systems to model kinematically, demanding high spatial and spectral resolution with excellent signal-to-noise. Only by assembling large, uniformly analyzed samples with well-defined selection functions can we robustly distinguish between different dwarf populations and uncover their formation pathways.

\subsection{Environmental modulation of the low-mass population}
\label{Sec:Environment}

The physics governing the dark and dim Universe---i.e., the low-mass end of galaxy formation---does not occur in isolation. At this extreme of the mass spectrum, even modest external forces---such as ram pressure, tidal interactions, or ionizing radiation---can outweigh internal feedback processes. Environment is therefore not a secondary influence but an intrinsic boundary condition for the evolution of dark and faint galaxies. Understanding these effects is essential for interpreting the cosmological statistics of the low-mass Universe.

Concerning primordial dark RELHICs, environment plays a decisive role in shaping their observable properties. These systems inhabit shallow potential wells that can barely retain their baryons once exposed to the heating from cosmic reionization. Consequently, their loosely bound gas is highly vulnerable to external perturbations. In dense environments---such as the outskirts of massive galaxies or within galaxy groups---the ambient medium can efficiently strip this gas through ram-pressure interactions, rapidly dispersing the neutral component and leaving the systems effectively gasless~\citep[e.g.,][]{Benitez-Llambay2017}. Conversely, RELHICs that evolve in relative isolation are expected to preserve their neutral gas, allowing them to remain detectable as compact \hi\ clouds. Quantifying how environmental processes regulate this balance is crucial for understanding the demographics of gas-rich low-mass halos and for constraining the interplay between reionization and structure formation on the smallest scales. Achieving this requires a combined theoretical and observational approach.  

Similarly, low-mass dwarf galaxies---which dominate the faint end of the galaxy mass function---are highly sensitive to both internal feedback and external environmental influences. The interplay between these processes leaves measurable imprints on the spatial distribution of their gas, and consequently, their \hi\ content~\citep[and references therein]{Benitez-Llambay2013, Thompson2023, Herzog2023}. Observationally, for example, there is ample evidence that low-mass satellites of more massive host galaxies tend to be gas-poor, while their counterparts in the field are gas-rich (e.g. \citealt{Grcevich2009,Spekkens2014,Putman2021,Geha2024}). By examining how the \hi\ distribution in low-mass galaxies varies with mass and environment, forthcoming SKA observations will offer an empirical manner to disentangle the relative roles of internal feedback and environmental interactions in shaping the neutral gas reservoirs of faint galaxies.

Moreover, statistical studies of low-mass dwarf galaxies in \textit{void} regions is particularly exciting.  These galaxies should be relatively unperturbed relative to those in more dense environments and highlight the effects of internal feedback on galaxy structure.  Studies have consistently shown that void galaxies tend to be gas rich \citep{Kreckel2012,Chengalur2013,Pustilnik2016a}.  Moreove, they also tend to be metal poor \citep{Pustilnik2016b}.  All of these signs point to void galaxies being relatively `pristine', suggesting that they excellent targets for studies of galaxy evolution \citep{Etka2008,Kreckel2011,Kurapati2024}.  With the SKA-Mid at AA4, it will be possible to probe these down to lower masses at large enough numbers to understand the effect of feedback and secular processes on low-mass galaxies on a statistical basis.  Such studies will, in turn, complement and enhance our understanding of galaxy structure and growth. 

Among scaling relations, the \hi size-mass relation is one of the tightest, extending over many orders of magnitude \citep{Wang2016}, and is apparently insensitive to environment~\citep[e.g.,][]{Wang2016, rajohnson2022, holwerda2011, Koribalski2018}. However, the same cannot be said for the relationship between the \hi\ size and the stellar size.
%Deviations from this relation at the faint end therefore provide a sensitive probe of how feedback and external forces reshape the gas reservoirs of low-mass systems. Recent deep observations are beginning to reveal these deviations. 
In the Fornax cluster, for instance, gas-rich dwarfs show systematically lower \hi\ surface densities than nearly isolated counterparts of similar mass, by factors of three or more~\citep{chamba2024a, chamba2024b}. As a result, the canonical 1\,$\rm M_{\odot}$\,pc$^{-2}$ isocontour that defines the \hi\ radius often lies within the stellar truncation radius, leading to smaller \hi–to–stellar size ratios. This pattern implies that ram pressure and tidal stripping remove the outer, low-density gas that otherwise maintains the tight relation observed in more massive galaxies~\citep{kleiner2023, chamba2024a}. Similar effects are emerging in other cluster and group environments~\citep[e.g.,][]{2023PASA...40...17W, zhu2024, blana2025}. %These results highlight that the apparent universality of the \hi\ size–mass relation breaks down at the faint end, precisely where the environment and internal feedback exert their strongest influence.

This issue connects to a long-standing concern in the optical regime, where half-light radii or isophotal diameters can miss the faint outskirts of galaxies~\citep{trujillo2020}. In the \hi\ domain, the situation is analogous: the canonical \hi\ radius may not trace the full gas extent in dwarfs, especially when external processes lower their overall \hi\ surface densities. Consequently, scaling relations derived from this `isocontour' may obscure real physical variations linked to environment or feedback~\citep[see also][]{reynolds2023}. This limitation is particularly relevant for the interpretation of ultra-diffuse and other dim galaxies, whose gas and stellar components often truncate at different radii~\citep[see][and Sec.~\ref{Sec:bTFR}]{chamba2020}.

In nearly isolated low-mass galaxies, a wide range of HI–to–stellar size ratios is observed, pointing to the complementary role of internal processes. Stellar feedback can drive inflows and outflows that redistribute gas and alter its surface-density profile~\citep[e.g.,][]{elbadry2016, rey2022, jones2025}. Depending on the recent star formation history and feedback efficiency, the canonical \hi\ radius may lie either within or beyond the stellar boundary~\citep{chamba2024a, chamba2024b}. Thus, both environment and feedback modulate how galaxies populate the \hi\ size–mass plane, even if the mean relation remains approximately intact for higher-mass systems.

Despite numerous targeted surveys, only a few dozen dwarf galaxies with $M_{\rm \hi}<10^{8}\,\rm M_{\odot}$ have been resolved well enough to explore these effects in detail~\citep[see][]{Wang2016, rajohnson2022, reynolds2023, chamba2024b, OBeirne2025}. The SKA will dramatically expand this sample, delivering homogeneous measurements of \hi\ distributions across a wide range of environments. By probing lower \hi\ surface densities~\citep{maccagni2024} and combining its data with deep optical imaging from LSST, \textit{Euclid}, and \textit{Roman}, the SKA will enable systematic tests of how feedback and environment shape the dark and dim Universe. These joint datasets will allow direct comparisons of stellar and gaseous truncations for large samples of galaxies, providing a physically grounded framework to link dark, dim, and environmentally modulated systems across cosmic time. In this way, SKA observations will bridge cosmology and galaxy evolution, connecting the predicted abundance of low-mass halos in $\Lambda$CDM to the observable diversity of faint, gas-rich dwarfs in the nearby Universe.

\section{Prospects for the SKA at AA4}
\label{sec:Prospects}

Addressing the core science questions articulated above requires more than just the high sensitivity of a single observatory; it demands a coordinated, multi-wavelength strategy. This section outlines the prospects for tackling these questions with SKA-Mid at AA4, beginning with the key synergies that will enable such progress.

Unlocking the full potential of the SKA will require other observatories to provide a full, multi-wavelength view of the low-mass Universe.  Together, these observatories will transform the landscape of the dark and dim Universe. Some of these facilities will begin operations before AA4 and, as such, will pave the way for the transformative capabilities of the SKA.  Given their importance, this section begins with a brief discussion of upcoming facilities before moving on to the specific predictions for SKA-Mid at AA4.

%Before discussing the specific predictions and capabilities of SKA-Mid at AA4, it is worth briefly considering the complementary roles of other current and forthcoming facilities that will shape the landscape of low-mass galaxy studies in the coming decade. Together, these observatories will both prepare the ground for, and later operate synergistically with, the transformative capabilities of the SKA.

\subsection{Synergy with Other Observatories}
\label{sec:OtherObservatories}

There are several currently operating and upcoming facilities that are relevant to the study of the dark and dim Universe.  These include radio observatories like FAST, CHORD, and the DSA-2000, as well as optical facilities like the Rubin Observatory and the Extremely Large Telescope (ELT).  Additionally, a number of critical space telescopes that are already transforming our understanding of galaxy formation and evolution.

FAST, currently the largest single-dish radio telescope in operation, will be crucial for answering the core science questions posed above. FAST has already demonstrated its power in the low-mass regime, contributing to new determinations of the \hi\ mass function (HIMF) and detecting numerous dark-galaxy candidates~\citep[e.g.,][]{Zhang2024, Ma2025}. As its survey programs mature, FAST is expected to produce the deepest HIMF yet measured and to compile extensive catalogs of candidate dark systems and low-mass dwarfs. While its Northern Hemisphere location limits direct sky overlap, it provides an essential large-area statistical comparison to the Southern Hemisphere surveys from SKA-Mid, testing the universality of the results. Additionally, it is expected that FAST, but more importantly the SKA's precursor MeerKAT, provide an invaluable pool of sources for early high-resolution follow-up. Early SKA-Mid science operations will therefore benefit greatly from this synergy, enabling mapping and kinematic studies of known MeerKAT detections, as well as FAST detections in the partial overlap region.

Two additional next-generation radio facilities, CHORD~\citep{Vanderlinde2019} and DSA-2000~\citep{Hallinan2019}, will further extend this progress. Both are wide-field interferometers located in the Northern Hemisphere and are optimized for large-area surveys. With sensitivities reaching \hi\ masses of $M_{\rm HI} \sim10^{5}-10^{6}\,\mathrm{M_\odot}$, they will be able to probe the faint end of the HIMF in the field and at moderate redshifts, tracing the evolution of neutral gas across cosmic time. These surveys will provide a vital statistical comparison sample for SKA studies of the same mass range.

Moving to the optical domain, the \textit{Vera C. Rubin Observatory’s Legacy Survey of Space and Time} (LSST) will play a decisive role in mapping the dim galaxy population and ruling out faint stellar counterparts of RELHIC candidates. Over the next decade, LSST is expected to reveal the majority of nearby %($d\lesssim $?? Mpc) 
low-mass and low-surface-brightness galaxies across the sky. %(with $M_{*} \gtrsim $ ?? $\rm M_{\odot}$). 
The completition of this survey will coincide with the commencement of SKA-Mid Key Science Projects, ensuring deep optical coverage and precise stellar characterization for nearly all detectable, non-dark systems. These data will provide essential targets for SKA-Mid follow-up and enable joint analyses of stellar and gaseous components at the faint end and will help pin point the best nearby RELHIC candidates for quick follow-up.

Accurate distance measurements are equally crucial for interpreting the \hi\ detections from SKA-Mid. The array’s sensitivity at AA4 will resolve $10^{7}\,\rm M_\odot$ systems out to $\sim10$~Mpc by more than 5 beams (using a $4.3''$ beam) and $10^{8}\, \rm M_\odot$ systems out to $\sim50$~Mpc (see Section~\ref{sect:size_distance_sensitivity}). Within the local volume precise distance measurements are essential (at larger distances Hubble-flow distances will suffice for mass estimates).  Forecasts suggest that the \textit{James Webb Space Telescope} can obtain tip-of-the-red-giant-branch (tRGB) distances to $\sim$50~Mpc~\citep{Newman2024}. Although JWST may no longer be operational by the time SKA-Mid reaches AA4, comparable capabilities are expected from the \textit{Roman Space Telescope} and the ground-based \textit{Extremely Large Telescope} (ELTs), ensuring accurate distances for most galaxies with stellar counterparts. The faintest, truly dark candidates, however, will continue to challenge even these facilities.

Together, these observatories will establish a rich, multi-wavelength framework within which SKA-Mid can operate while maximizing the science returns. They will provide the critical context, target lists, and comparative baselines that will make SKA observations of the dark and dim Universe both efficient and interpretable.

\subsection{Estimates for SKA-Mid at AA4}
\label{sect:size_distance_sensitivity}

The science goals related to the dark and dim Universe demand a careful balance between angular resolution and sensitivity. The size of an \hi\ disk, $R_{\rm HI}$, is generally defined as the radius where the surface density reaches $1\,\rm M_{\odot}\,pc^{-2}$~\citep{Wang2016}. This definition is physically motivated and observationally useful, as there is a well-established correlation between \hi\ size and total \hi\ mass that spans seven orders of magnitude. To distinguish between different types of dark candidates, identify morphological disturbances, or infer dynamical structure, at more than $1.5$ full beams across the disk are required, while kinematic modeling of the DM content typically demands at least five beams across the \hi\ extent~\citep{Deg2022}.

\begin{figure}[h]
    \centering
	\includegraphics[width=0.9\columnwidth]{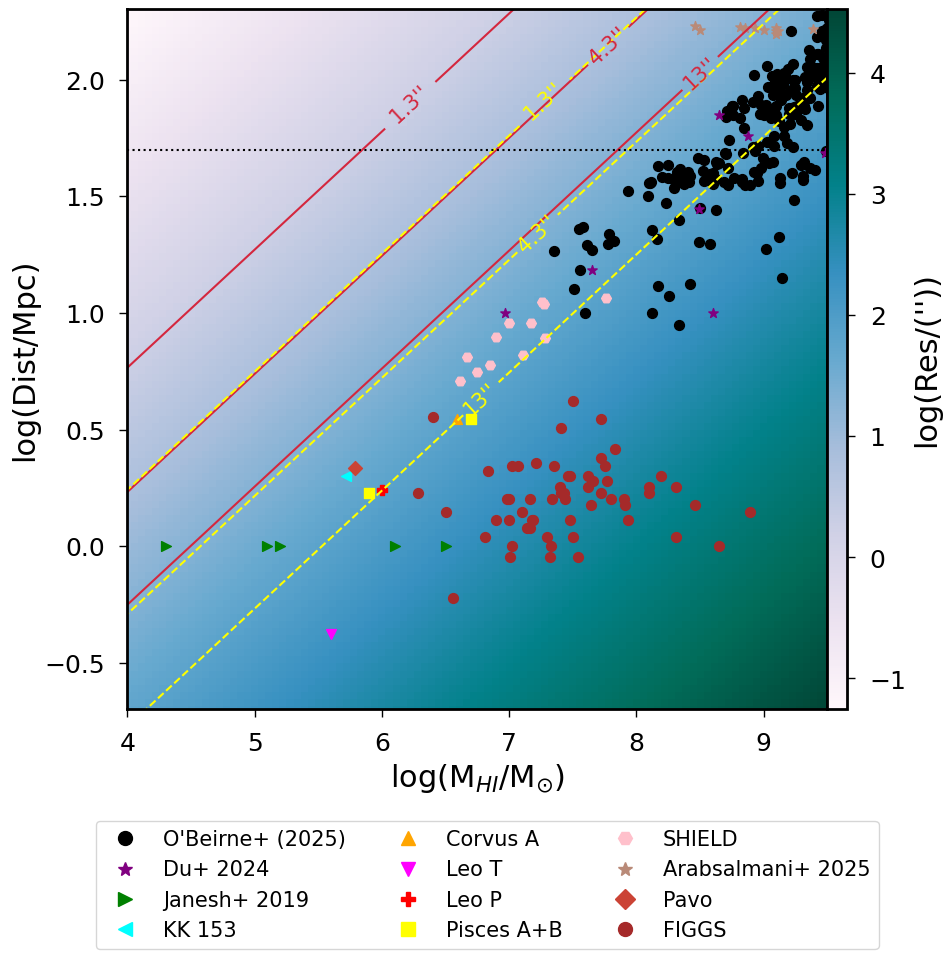}
    \caption{The expected distances to galaxies that can be resolved by various AA4 fiducial surveys.  The solid red and dashed yellow lines indicate constant observed angular resolutions inferred by the \citet{Wang2016} size-mass relation.  The solid red lines are for 1.5 beams and the dashed yellow lines are for 5 beams.  Each line is labelled by the specific AA4 resolution (see Table \ref{tab:HIMFPredictions} for the mock survey details).  The various colored points indicate the distance and mass for particular objects and surveys.  For clarity, they do not indicate the specific angular size of these objects.  They generally follow diagonals due to the sensitivity of the observatories used to detect the \hi\ in each of the galaxies. The dashed horizontal line is at 50 Mpc and indicates the distance out to which the JWST can potentially measure tRGB distances \citep{Newman2024}.}
    \label{fig:SizeDistanceDist}
\end{figure}

Taking these requirements into account, Figure~\ref{fig:SizeDistanceDist} illustrates the approximate distances out to which galaxies of given \hi\ masses can be resolved by 1.5 beams (solid lines) and 5 beams (dashed lines) for a number of AA4 resolutions (based on the SKA-Mid sensitivity calculater using Briggs +1 tapering). The physical size of a galaxy with a given \hi\ mass is estimated from the \citet{Wang2016} size--mass relation,
\begin{equation}
    \log(D_{\rm HI}) = 0.506 \log(M_{\rm HI}) - 3.293,
\end{equation}
where $D_{\rm HI} = 2R_{\rm HI}$. This size can then be converted into an angular resolution at a particular distance, producing the color-coded map shown in Figure~\ref{fig:SizeDistanceDist}. Lines of constant resolution thus appear as diagonals on this plane.

For comparison, Figure \ref{fig:SizeDistanceDist} shows the measured \hi\ mass and distances to a number of different low-mass samples.  It includes the WALLABY pilot low surface brightness systems found in \citet{OBeirne2024}, the low-mass SHEILD \citep{Cannon2011,McQuinn2014} and FIGGS \citep{Begum2008a}samples. It also includes some specific local ultrafaint dwarfs with known \hi\ masses; Leo P \citep{Giovanelli2013,BernsteinCooper2014}, Leo T \citep{Irwin2007,Adams2018}, Pisces A and B \citep{Tollerud2015}, Corvus A \citep{Jones2024}, Pavo \citep{jones2025},and KK 153 \citep{Yu2025}. Additionally it includes the faint galaxies of \citet{Du2024}, the ultra-faint galaxies of \citet{Janesh2019}.  The final set of galaxies are the dark galaxy candidates of \citet{Arabsalmani2025}.  It is worth noting that these dark galaxy candidates are not RELHICS, having masses closer to $10^{9}~\msol$.  They are thought to form near cosmic filaments due to DM overdensities, leading to their colloquial name of `black pearl' galaxies. 

The various low-mass galaxies shown in Figure \ref{fig:SizeDistanceDist} are their to highlight known samples of dark and dim galaxies, but they are not exhaustive.  Their location in the figure is based purely on their reported distances and masses, but, if they do follow the size-mass relation, it also provides a rough estimate of their observed resolution.  These galaxies typically follow diagonals due to the sensitivities of the observatories used to detect them.  Assuming that to be true, almost of all these galaxies would be observed by $\ge5$ beams of resolution with a $4.3''$ beam, which would enable kinematic modelling.  Moreover, some targeted very low-mass local group galaxies could be modelled by a $1.3''$ beam. 

Considering the larger picture presented by Figure \ref{fig:SizeDistanceDist}, it will be possible to model galaxies down to $10^{6}~\msol$ out to a distance of $\sim5$ Mpc, and $10^{7}~\msol$ out to $\sim15$ Mpc using a $4.3''$ beam.  By contrast a $1.3''$ beam will enable modelling at $10^{7}~\msol$ out to 50 Mpc.  This will provide a sample of kinematic models that can fill in the low-mass end of the bTFR as well as the low rotation end of the HIVF.  

\begin{table}[]
    \centering
    \begin{tabular}{|c|c|c|c|c|c|c|c|c|}
      \hline
      Resolution   & 6.0 - 6.5 & 6.5 - 7.0 & 7.0 - 7.5 & 7.5 - 8.0 & 8.0 - 8.5 & 8.5 - 9.0 & 9.0 - 9.5 & Total\\ 
      \hline 
      \hline
      \multicolumn{9}{|c|}{AA*} \\
      \hline 
      \multicolumn{9}{|c|}{$4.45''$ Beam, $100^{\circ^{2}}$ Area, 81 $\mathrm{\mu}$Jy Beam$^{-1}$ Noise, 3.1 $\kms$ channel,  12 hr exposure, 1200 hr Total}\\
      \hline
       Unresolved & $60$  & $2.4\times10^{2}$ & $9.3\times10^{2}$ & $3.4\times10^{3}$ & $1.1\times10^{4}$ & $3.4\times10^{4}$ & $8.4\times10^{4}$ & 1.3$\times10^{5}$ \\
       1.5 Beams  & 6  & 26 &$1.2\times10^{2}$& $5.6\times10^{2}$& $2.8\times10^{3}$& $1.6\times10^{4}$& $7.9\times10^{4}$& $9.7\times10^{4}$ \\
       5 Beams  & 0.2  & 0.7 & 3 & 13 & 58 & $2.6\times10^{2}$  & $1.1\times10^{3}$& $1.4\times10^{3}$\\
      \hline
      \multicolumn{9}{|c|}{$13.15''$ Beam, $1000^{\circ^{2}}$ Area, 447 $\mathrm{\mu}$Jy Beam$^{-1}$ Noise, 3.1 $\kms$ channel, 30 min exposure, 500 hr Total}\\
      \hline
       Unresolved & $49$  & $2.0\times10^{2}$ & $8.1\times10^{2}$ & $3.1\times10^{3}$ & $1.2\times10^{4}$ & $3.9\times10^{4}$ & $1.1\times10^{5}$ & 1.6$\times10^{5}$ \\
       1.5 beams   & 2  & 10 & 45 & $2.0\times10^{2}$ & $8.9\times10^{2}$ & $3.9\times10^{3}$ &$1.7\times10^{4}$ & $2.2\times10^{4}$\\
       5 beams  &  0 & 0.3 & 1 & 5 & 22  & 91 &$3.5\times10^{2}$ & $4.7\times10^{2}$\\ 
      \hline 
      \hline
      \multicolumn{9}{|c|}{AA4} \\
      \hline 
      \hline
      \multicolumn{9}{|c|}{$1.29''$ Beam, $1^{\circ^{2}}$ Area, 8.86 $\mathrm{\mu}$Jy Beam$^{-1}$ Noise, 3.1 $\kms$ channel, 500 hr exposure, 500 hr Total}\\
      \hline 
       Unresolved  & 14  & 54 & $1.9\times10^{2}$  & $5.9\times10^{2}$ &$1.7\times10^{3}$ &$4.2\times10^{3}$ & $8.9\times10^{3}$& $1.6\times10^{4}$ \\
       1.5 Beams  & 3  & 13 &  69 & $4.3\times10^{2}$& $1.7\times10^{3}$ &  $4.2\times10^{3}$ &$8.9\times10^{3}$ & $1.5\times10^{4}$ \\
       5 beams & 0  & 0.3 & 1 & 7 & 33 & $1.9\times10^{2}$  & $1.4\times10^{3}$ & $1.6\times10^{3}$ \\      
       \hline
       \multicolumn{9}{|c|}{$4.39''$ Beam, $100^{\circ^{2}}$ Area, 77.2 $\mathrm{\mu}$Jy Beam$^{-1}$ Noise, 3.1 $\kms$ channel,  8 hr exposure, 800 hr Total}\\
       \hline 
       Unresolved  & $65$  & $2.6\times10^{2}$ & $1.0\times10^{3}$  & $3.7\times10^{3}$ &$1.2\times10^{4}$ & $3.6\times10^{4}$&$8.9\times10^{4}$ & $1.4\times10^{5}$ \\
       1.5 beams   & 6  & 27 & $1.3\times10^{2}$ & $5.9\times10^{2}$& $2.9\times10^{3}$ & $1.6\times10^{4}$& $8.3\times10^{4}$ &$1.0\times10^{5}$ \\
        5 beams  & 0.2  & 0.7 & 3 & 14 & 61 & $2.7\times10^{2}$ & $1.1\times10^{3}$& $1.4\times10^{3}$\\
       \hline
       \multicolumn{9}{|c|}{$13.42''$ Beam, $1000^{\circ^{2}}$ Area, 491 $\mathrm{\mu}$Jy Beam$^{-1}$ Noise, 3.1 $\kms$ channel, 15 min exposure, 250 hr Total}\\
       \hline 
       Unresolved  &  42 & $1.8\times10^{2}$ & $7.0\times10^{2}$  & $2.8\times10^{3}$ &$1.0\times10^{4}$ & $3.5\times10^{4}$& $9.8\times10^{4}$& $1.5\times10^{5}$ \\
       1.5 beams & 2   &10  & 41 & $1.8\times10^{2}$&$8.0\times10^{2}$& $3.5\times10^{3}$&  $1.5\times10^{4}$&  $2.0\times10^{4}$ \\
       5 beams  & 0   &0.3  & 1 & 5 &20  & 82 & $3.2\times10^{2}$& $4.2\times10^{2}$\\
      \hline
      
    \end{tabular}
    \caption{Predictions for the number of detections as a function of mass for a variety of fiducial Band 2 surveys.  The top row lists the \hi mass for each column (i.e. the 6.0 - 6.5 column consists of predictions for the number of galaxies with \hi\ masses in the range $6.0 \le \log_{10}(M_{\hi}/\msol) \le 6.5)$.  The predictions are based on the \citet{Jones2018} HIMF coupled with the \citet{Wang2016} size-mass relation.  All surveys use Briggs + 1 taperings. For the resolved numbers, the sizes are based on the $1~\msol~\textrm{pc}^{-2}$ isodensity size (i.e. the size used for the \citet{Wang2016} size-mass relation).  To be detected, the object must be resolved by the number of beams and to be at least a 3 $\sigma$ detections over the noise at Z=0 over $16~\kms$.  At the upper limit of Band 2 (i.e. Z=0.5), the surface densities may fall below this $3~\sigma$ limit due to the $(1+Z)^{4}$ dimming.  The unresolved numbers are based on a $5~\sigma$ limit over $16~\kms$ with $13.4~\textrm{kHz}$ channels.}
    \label{tab:HIMFPredictions}
\end{table}

Resolution, however, comes at the cost of sensitivity. Achieving the fiducial surface density threshold of a $3~\sigma$ detection at $1\rm\,M_\odot\,\pc^{-2}$ across $16~\kms$ using $13~\textrm{kHz}$ channels requires progressively longer integrations at finer resolutions. For instance, the SKAO sensitivity calculator shows that a 13$''$ beam requires only $\sim$15 minutes of integration to reach this limit, whereas achieving the same depth at $4.3''$ takes roughly 8 hours, and at $1.3''$ would require nearly 500 hours. Consequently, addressing different science questions will required tailored observing strategies.

To illustrate this trade-off quantitatively, Table~\ref{tab:HIMFPredictions} lists the expected number of detections per mass bin for fiducial untargeted SKA-Mid Band 2 surveys for unresolved detections as well as at 1.5 and 5 beam resolutions.  The estimated numbers are based on drawing from the \citet{Jones2018} HIMF, with size limits based on the \citet{Wang2016} size–mass relation and sensitivity limits from the SKAO sensitivity estimator. The calculations account for the $(1+Z)^4$ cosmological dimming as well as the effect of redshift on the projected size.  They do not account for RFI or other effects.  Explicitly, the resolution and noise are determined using SKA-Mid options with a declination of $-45^{\circ}$, elevation of $60^{\circ}$, and with Briggs +1 tapering.  Ultimately, Table \ref{tab:HIMFPredictions} demonstrates that that science focusing on masses below $10^{6}\, \rm M_\odot$ will require targeted rather than untargetted surveys. Between $10^{6}$ and $10^{7}~\msol$, kinematic modelling studies will likely require targeted observations rather than drawing from an untargetted survey.  That being said, for low-mass studies, a resolution near $4.3''$ appears optimal, yielding the highest number of sub-$10^{8}~\msol$ detections. For calculations of the HIMF, it would take only $\sim$800 hours of SKA-Mid at AA4 observations to produce a comparable number of low-mass detections to the full ALFALFA survey~\citep{Jones2018}.  But, unlike ALFALFA, many of these detections, particularly above $10^{7.5}~\msol$ will be resolved.  Another key difference is that the low mass detections will be at larger distances than equivalent ALFALFA detections.  As such, the relative uncertainty in the distances will decrease (assuming that direct distance estimates are unavailable).  Returning to Figure \ref{fig:HIMF}, an untargeted, medium-deep Band 2 survey will resolve the low mass end of the HIMF, which will provide an anchor for studies of the HIMF in more dense environments.  

\begin{figure}[h]
    \centering
	\includegraphics[width=0.9\columnwidth]{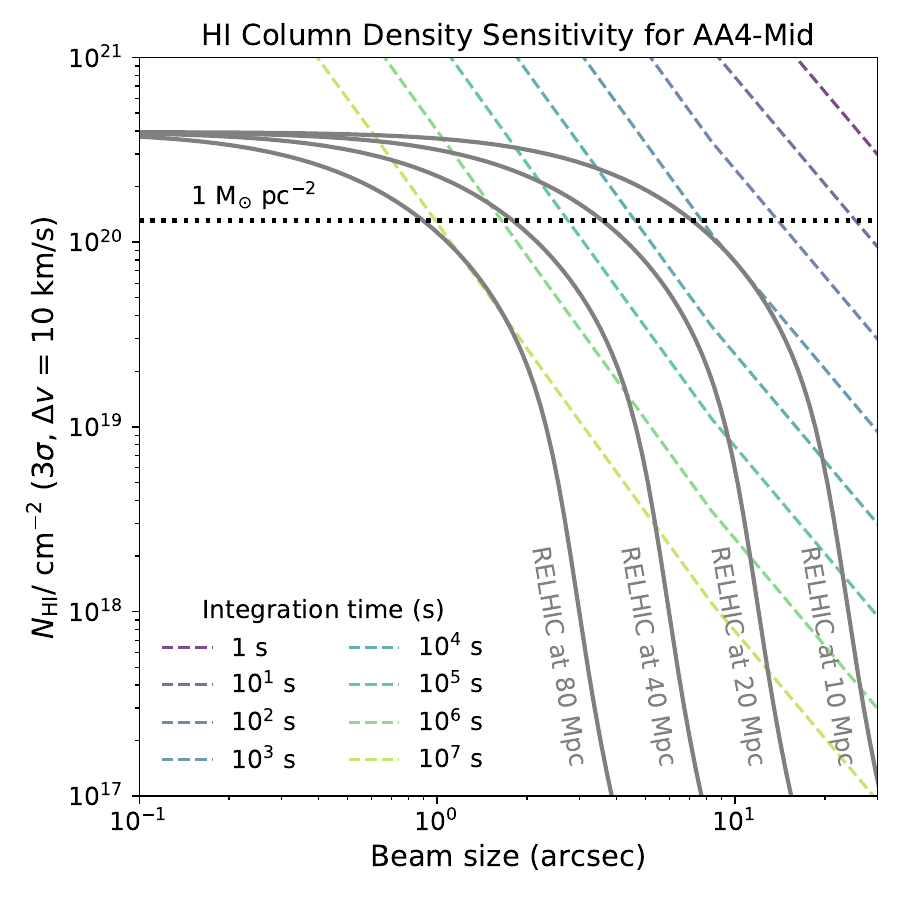}
    \caption{Different coloured lines indicate the $3\sigma$ column density detection limit for a velocity window $\Delta v = 10~\mathrm{km\,s^{-1}}$, shown as a function of beam size for various integration times, as indicated by the legend. The grey solid lines show the intrinsic column density profiles of RELHICs similar to Cloud-9, i.e. systems with masses, $M_{200} \approx 10^{9.7}\,M_{\odot}$, and $M_{\rm HI} \approx 10^6 \ M_{\odot}$, placed at different distances from 10~Mpc up to 80~Mpc. This suggests that targeted observations could resolve the detailed structure of RELHICs out to distances of 20--40~Mpc with integration times of a few hours. However, the small beam required to resolve more distant RELHICs would make such observations challenging, given the extremely long integration times needed to achieve sufficient sensitivity. These more distant RELHICs would be better targeted as unresolved systems using larger beams.}
    \label{fig:RELHICS-AA4}
\end{figure}

Since RELHICs do not necessarily follow the \hi\ scaling relations of dim galaxies, Figure~\ref{fig:RELHICS-AA4} illustrates the sensitivity of the AA4-Mid array for resolving and mapping the neutral-hydrogen content of RELHICs. As before, the data were generated using the SKA sensitivity calculator for the AA4-Mid configuration in Band 2, but assuming now target observations at a declination of -45 degrees and an elevation of 45 degrees, centered at a frequency of 1420 MHz. The setup uses a continuum bandwidth of 200 MHz, unity spectral averaging, and uniform image weighting. The different points on the sensitivity curves are achieved by modifying the beam size via tapering in the calculator.

The figure displays the $3\sigma$ column density detection limit as a function of beam size. Column densities are derived from the surface-brightness temperature, $T_{\rm b}({\nu})$ through the standard equation relating the surface brightness temperature and column density for optically thin gas, for a velocity window of $\Delta v=10~\text{km s}^{-1}$~\citep[e.g.][]{Dickey1990}.\footnote{$N_{\rm HI} \approx 1.823 \times 10^{18} \ {\rm cm^{-2}} \displaystyle\int{}{} T_{\rm b} (v) dv$.} The different coloured lines represent various integration times, as indicated in the legend. These sensitivity limits are compared against the intrinsic column density profiles of RELHIC models (grey solid lines) similar to Cloud-9, with total DM and \hi\ masses, $M_{200} \approx 10^{9.7}~M_{\odot}$ and $M_{\rm HI}\approx 10^{6} \ M_{\odot}$, respectively, placed at increasing distances from 10~Mpc to 80~Mpc.

Figure~\ref{fig:RELHICS-AA4} demonstrates the transformative power of AA4 in our ability to probe the fundamental nature of DM. For a nearby RELHIC candidate (e.g., at $\sim$10 Mpc), SKA-Mid in its AA4 configuration will achieve the required sensitivity to map its neutral gas distribution with a resolution of $\sim$10 arcsec in a matter of hours. This capability allows for a robust characterization of the overall gas morphology and kinematics. More critically, for primary targets, relatively brief integrations can reach $\sim4''$ resolutions (see also Table \ref{tab:HIMFPredictions}), while deep integrations of $\sim100$ hours can potentially reach $\sim2''$ or lower, enabling us to resolve the gas structure within the central kpc---the very region where competing DM models make their most divergent predictions. RELHICs, being starless, offer the only known laboratories where the primordial DM density profile can be measured from their gas distribution free from baryonic contamination.

Complementary to this, a targeted, high resolution study of a number of low-mass, dim galaxies will be valuable (with resolution $\le 4''$ and matched to that in RELHIC studies). The high resolution coupled with 3D kinematic modelling will remove many of the systematic effects related to beam smearing that cause issues in prior studies of the core-cusp problem.  This will enable a measurement of the inner DM profiles of low mass in the presence of baryons.  By comparing these measurements to those from RELHICs, it will be possible to directly constrain the effect of baryonic feedback processes in some of the lowest mass galaxies in the local Universe.

Thus, SKA-Mid's ability to resolve the inner structure of these nearby clouds is not merely an incremental improvement. It brings the cusp-versus-core problem from a longstanding, model-dependent debate (see Sec.~\ref{sec:Introduction}) into a directly answerable observational question. While Figure~\ref{fig:RELHICS-AA4} confirms that resolving the inner structure of more distant RELHICs (> 40 Mpc) will remain observationally prohibitive, these systems will become prime targets for large-area untargetted surveys. By detecting them as unresolved sources with larger beams to maximize survey speed, SKA-Mid will build the first statistically significant census of the starless halo population, providing the crucial complementary test of their predicted abundance (see Figure~\ref{fig:HIMF}).

\section{Final Summary and Outlook}
\label{sec:Summary}

This chapter has explored the low-mass frontier of galaxy formation, a regime defined by the tension between the predicted abundance of DM halos in the $\Lambda$CDM paradigm and the observed census of faint galaxies (Sec.~\ref{sec:Introduction}). These discrepancies imply that baryonic processes must strongly suppress galaxy formation in shallow potential wells, sculpting the cosmic landscape into two distinct but physically linked populations: a vast number of starless ``dark'' halos (RELHICs) and a crucial population of faint, gas-rich ``dim'' systems. Together, they represent a critical testbed for both cosmology and the physics of galaxy formation and evolution.

We have detailed the core science questions arising from this cosmological framework (see Sec.~\ref{sec:CoreScience}). The search for the dark Universe is primarily an effort to confirm the existence of a cosmologically significant population of primordial, starless Reionization-Limited \hi\ Clouds (RELHICs) (see Sec.~\ref{Sec:RELHICs}). These systems offer an unparalleled laboratory for probing the pristine inner density structure of DM halos, providing a clean test of the cusp-core problem, free from the baryonic feedback that complicates its interpretation in luminous galaxies. A formidable observational challenge, however, is to unambiguosly distinguish these primordial objects from astrophysical contaminants such as tidal debris and pressure-confined clouds (see Sec.~\ref{Sec:RELHICs}). 

In parallel, and complementing this direct search is the statistical study of the dim Universe (see Sec.~\ref{sec:DimUniverse}). Through robust measurements of the HIMF (see Sec.~\ref{Sec:HIMassFunction}), the HIVF (see Sec.~\ref{Sec:HIVelFunction}), and the bTFR (see Sec.~\ref{Sec:bTFR}), the faint galaxy population provides powerful, integrated constraints on the efficiency of gas suppression, the abundance of low-velocity systems, and the predicted connection between galaxies and their host halos. Across both populations, we have underscored that environment is not a secondary influence but an intrinsic and often dominant player that modulates the gas content, morphology, and ultimate fate of these fragile low-mass systems (see Sec.~\ref{Sec:Environment}).

Finally, we outlined the transformative prospects for addressing these questions with the Square Kilometer Array (see Sect.~\ref{sec:Prospects}), emphasizing that progress requires a coordinated, multi-wavelength strategy. The SKA will operate in synergy with pathfinder facilities and existing observatories like FAST---which provide catalogs of candidate objects--- and next-generation optical surveys such as LSST---which will deliver the essential characterization of the optical counterparts needed to identify both dim galaxies and RELHICs (see Sec.~\ref{sec:OtherObservatories}). 

The capabilities of the SKA-Mid array at AA4, as quantified in this chapter (see Sec.~\ref{sect:size_distance_sensitivity}), represent a paradigm shift. Its sensitivity and angular resolution will, for the first time, enable observers to resolve the inner kiloparsec of nearby RELHIC candidates (see Sec.~\ref{Sec:RELHICs}). This capability will move the cusp-core debate from a model-dependent problem to a direct observational question. Simulateneously, its deep, wide-area surveys will build the statistically significant and kinematically modeled samples of dim galaxies required to definitively map their statistical distributions across cosmic environments (see Sec.~\ref{sect:size_distance_sensitivity}). By illuminating both the individual dark halos and the collective dim population, the SKA is poised to provide a complete and physically grounded understanding of the interplay between cosmology and galaxy formation at the smallest cosmic scales. 

\section*{Acknowledgements}
The authors wish to thank the anonymous referee for their useful comments.  The authors further wish to thank M. Jones, A. Kazemi-Moridani, D. Kleiner, and A. Ponomareva for their generous sharing of observed HIMF measurements.  They wish to thank J. Mayor for their simulated HIMF measurements.  The authors also wish to thank M. Arabsalmani, N. Arora, M. Chhabra, and L. Chemin for useful discussions in the preparation of this chapter.

ABL acknowledges support by the Italian Ministry for Universities (MUR) program “Dipartimenti di Eccellenza 2023-2027” within the Centro Bicocca di Cosmologia Quantitativa (BiCoQ), and support by UNIMIB’s Fondo Di Ateneo Quota Competitiva (project 2024-ATEQC-0050).

FMM carried out part of the research activities described in this paper with contribution of the Next Generation EU funds within the National Recovery and Resilience Plan (PNRR), Mission 4 - Education and Research, Component 2 - From Research to Business (M4C2), Investment Line 3.1 - Strengthening and creation of Research Infrastructures, Project IR0000034 – “STILES - Strengthening the Italian Leadership in ELT and SKA.

\bibliographystyle{abbrvnat-maxbibnames4}
\bibliography{chapter} % if your bibtex file is called example.bib

\end{document}